%% file: Main.tex
\documentclass[final,3p,12pt]{elsarticle}
\usepackage{lineno}
\usepackage{subfigure}

\usepackage{lipsum}

\usepackage{hyperref}
\usepackage[english]{babel}
\usepackage{amsmath}
\usepackage{graphicx}
\usepackage{gensymb}
\usepackage{soul}
\usepackage{lscape} 
\usepackage{verbatim} 
\usepackage{amssymb}
\usepackage{tablefootnote}
\usepackage{threeparttable}
\usepackage{footnote}
\usepackage{url}

\usepackage{array}
\usepackage{ragged2e}
\newcolumntype{P}[1]{>{\RaggedRight\hspace{0pt}}p{#1}}

\newcommand{\checkbox}[1]{%
  \ifnum#1=1
    \makebox[0pt][l]{\raisebox{0.15ex}{\hspace{0.1em}$\checkmark$}}%
  \fi
  $\square$%
}

\usepackage{array,booktabs} 
\usepackage{longtable}
\usepackage{supertabular}
\usepackage[justification=centering]{caption}
\newcolumntype{P}[1]{>{\centering\arraybackslash}p{#1}}
\usepackage{multirow}
\usepackage{xcolor}

\makeatletter
\def\ps@pprintTitle{%
  \let\@oddhead\@empty
  \let\@evenhead\@empty
  \def\@oddfoot{\reset@font\hfil\thepage\hfil}
  \let\@evenfoot\@oddfoot
}
\makeatother

\begin{document}





\title{Beyond Data, Towards Sustainability: A Sydney Case Study on Urban Digital Twins}

\author[add1]{Ammar Sohail}
\ead{ammar.sohail@monash.edu}

\author[add1]{Bojie Shen}
\ead{bojie.shen@monash.edu}

\author[add1]{Muhammad Aamir Cheema}
\ead{aamir.cheema@monash.edu}

\author[add2]{Mohammed Eunus Ali}
\ead{eunus@cse.buet.ac.bd}

\author[add3]{Anwaar Ulhaq}
\ead{a.anwaarulhaq@cqu.edu.au}

\author[add4]{Muhammad Ali Babar}
\ead{ali.babar@adelaide.edu.au}

\author[add1]{Asama Qureshi}
\ead{asama.qureshi@gmail.com}

\address[add1]{%
  Faculty of Information Technology, Monash University, Australia\\
  \textit{E-mail:} \{ammar.sohail, bojie.shen, aamir.cheema\}@monash.edu, asama.qureshi@gmail.com
}

\address[add2]{%
  Department of Computer Science and Engineering, BUET, Bangladesh\\
  \textit{E-mail:} eunus@cse.buet.ac.bd
}

\address[add3]{%
  School of Engineering and Technology, Central Queensland University, Australia\\
  \textit{E-mail:} a.anwaarulhaq@cqu.edu.au
}

\address[add4]{%
  Centre for Research on Engineering Software Technologies, University of Adelaide, Australia\\
  \textit{E-mail:} ali.babar@adelaide.edu.au
}

\begin{abstract}
As urban areas grapple with unprecedented challenges stemming from population growth and climate change, the emergence of urban digital twins offers a promising solution. This paper presents a case study focusing on Sydney's urban digital twin, a virtual replica integrating diverse real-time and historical data, including weather, crime, emissions, and traffic. Through advanced visualization and data analysis techniques, the study explores some applications of this digital twin in urban sustainability, such as spatial ranking of suburbs and automatic identification of correlations between variables. Additionally, the research delves into predictive modeling, employing machine learning to forecast traffic crash risks using environmental data, showcasing the potential for proactive interventions. The contributions of this work lie in the comprehensive exploration of a city-scale digital twin for sustainable urban planning, offering a multifaceted approach to data-driven decision-making.
\end{abstract}





\maketitle

\hspace{5mm}
\input{Introduction}

\input{Related_Work}
\input{Spatial_Digital_Twin}

\input{Techniques}

\section{Conclusions} \label{sec:conclusion}
In this paper, we design an urban digital twin that integrates multifaceted data from diverse sources, including air quality, weather, GHG emissions, and more. Our urban digital twin serves as a case study exploring several applications in urban sustainability. We present visualization tools that empower researchers and urban planners to assess suburb rankings based on their specific interests and to automatically extract statistical and spatial insights for in-depth data analysis. Furthermore, we demonstrate that the environmental data integrated into our urban digital twin can be effectively leveraged with machine learning techniques. Specifically, we showcase its application in predicting traffic crash risks considering environmental data, underscoring the practical implications of our approach for enhancing urban safety.


\bibliographystyle{cas-model2-names}


\bibliography{Digital_Twin}

\end{document}

%% file: Introduction.tex
\section{Introduction}

Urban areas around the world are facing unprecedented challenges due to rapid population growth and climate change. The complexity of modern cities demands innovative solutions that integrate vast and diverse datasets to inform decision-making processes. In this context, the concept of \emph{urban digital twin} or \emph{spatial digital twin}~\cite{dembski2020urban,ali2023enabling} has emerged as a powerful tool, enabling researchers and policymakers to create virtual replicas of physical environments. These digital twins, when populated with real-time and historical data, provide invaluable insights into urban dynamics, thereby guiding the formulation of sustainable policies and interventions.

There is a substantial body of research on urban digital twins, and many of the existing works center on mobility digital twins~\hbox{\citep{MDT,MDT2}} for managing entities like humans and vehicles, or urban digital twins~\hbox{\citep{SCDT_benifit, SCDT, SCDT2}} designed to enhance decision-making and urban planning. Additionally, there exists several urban digital twins of different cities or countries\hbox{~\citep{ali2023enabling}}, such as Singapore's digital twin emphasizing virtual experiments for network coverage analysis. Different cities, like Zurich and Boston~\citep{switerlandDT,bostonDT}, deploy digital twins for urban design scenarios and analytical tools for assessing building shadow projections, respectively. Moreover, New South Wales (NSW) digital twin concentrates on asset monitoring and simulating natural disasters. 

In this paper, we use Sydney as a case study and create its urban digital twin exploring some of its applications in urban sustainability. This digital twin  incorporates data from multifaceted sources, including weather records, crime statistics, greenhouse gas emissions, traffic volume, air quality measurements, and detailed records of traffic incidents. Much of this data is dynamic and our automated script fetches this data at the beginning of each day. By integrating these disparate datasets, we construct a comprehensive and intricate digital representation of Sydney, capturing the potential interactions between social, environmental, and infrastructural factors.
Through this case study, we hope to demonstrate how a data-driven approach can inform sustainable urban planning and decision-making.

We explore several applications of this digital twin for sustainability. 
We develop a visualization system tailored for this urban digital twin that does not only present the data in a visually engaging manner but also allows for spatial analysis. For instance, our system enables the ranking of Sydney suburbs based on specific parameters, such as emissions or traffic incidents, spanning different years. This spatial lens does not only enrich our understanding of the data but also provides a unique perspective for urban planners and researchers.

In addition to spatial visualization, we present techniques for automatically extracting insights from the data. One of our key objectives is to automatically identify intriguing correlations between various variables within the dataset. For instance, our system unveils the relationships between traffic incidents and air pollution levels, shedding light on potential causative factors and facilitating targeted interventions. Furthermore, our research delves into spatial correlation analysis, uncovering nuanced patterns that underlie the spatial distribution of different urban phenomena. For example, our system explores spatial clusters of crimes and wastewater emissions between neighborhoods, 
demonstrating that high wastewater emissions indicate poor infrastructure, which often leads to higher crime rates. This enables urban planners to take strategic actions and improve planning efforts through a thorough understanding of spatial relationships.

Finally, we investigate the potential applications of machine learning techniques for urban sustainability. 
Leveraging our urban digital twin, we deploy various machine learning methods to predict traffic crash risks considering the environmental data. The experimental results are promising considering different evaluation metrics, signifying a potential avenue for employing data-driven approaches to issue alerts for high-risk environmental conditions for traffic.
These predictions  open avenues for proactive interventions, enabling authorities to allocate resources effectively and implement preventive measures to reduce traffic incidents.

To summarize, we make the following contributions in this work.

\begin{itemize}
    \item We present a case study of an urban digital twin of Sydney that integrates diverse data from multiple sources and explore its applications in urban sustainability.
    \item Our system offers a range of visualization tools enhanced with advanced automatic data analysis techniques. It can mine statistical and spatial insights.
    \item We showcase how data-driven approaches can be effectively employed to predict traffic risks considering environmental data.
\end{itemize}

The rest of the paper is organized as follows. The existing literature is discussed in Section~\ref{Sec:Related_Work}. The details of our urban digital twin are presented in Section~\ref{sec:Digital_Twin}. Some representative sustainability applications of our urban digital twin are covered in Section~\ref{sec:applications}. We conclude the paper in Section~\ref{sec:conclusion}.

%% file: Related_Work.tex
\section{Literature Review}\label{Sec:Related_Work}
The concept and model of digital twin have been known for over two decades~\citep{grieves2002completing}, and have been evolved into various forms, including spatial digital twins, human digital twins, and more. 
In this section, we briefly introduce the digital twin and its variants. Subsequently, we delve into the details of spatial digital twins, also often known as urban digital twins, and summarize several case studies.

\subsection{Digital Twins and Variants}
The essential components of digital twins comprise three main elements: a physical object or process, a virtual representation of the object or process, and communication channels to exchange the information between physical and virtual representation. 
 Within this framework, \emph{industrial digital twins} concentrate on manufacturing automation, demonstrating adaptability in optimizing production processes and engineering product-family design~\citep{IDT_p, IDT_pf,umair2021impact}. Meanwhile, \emph{human digital twins} model the human life cycle, leveraging augmented reality for smart health management and enhancing personalized healthcare~\citep{HDT, HDP_2}. Beyond industry and health, digital twins find application in education, capturing classroom gaze behavior~\citep{classroom_dt}, and in the energy sector, integrating techniques for monitoring consumption and identifying challenges~\citep{energy_dt}. 
 Recognizing the significance of spatial data, a recent report~\hbox{\citep{SDT_WGIC}} underscores the importance of incorporating spatial data into digital twins, benefiting various applications across industries and the research communities.
 In the subsequent section, we explore spatial digital twins and highlight relevant case studies.

\subsection{Spatial Digital Twins and Case Studies}
Spatial digital twins, unlike traditional digital twins, incorporate spatial data to offer location-based representations of geo-spatial objects or processes. 
 \emph{Mobility digital twins} \citep{MDT} focus on managing entities like humans, vehicles, and traffic in a spatial region. 
A comprehensive framework presented by \citet{MDT} addresses these complexities, while \citet{MDT2} propose an enhanced mobility prediction system.
\emph{Urban digital twins}~\citep{SCDT_benifit} expand digital twins techniques into smart cities, benefiting stakeholders like city planners, policy-makers, and decision-makers.
\citet{SCDT} introduce a framework integrating Building Information Modeling (BIM) and Geographic Information System (GIS) data for urban digital twins,
while \citet{SCDT2} explores constructing urban digital twins through 3D point cloud processing.
Comprehensive surveys are also available, highlighting the current challenges and potential research opportunities for urban digital twins~\citep{SCDT_survey1, SCDT_survey2,ali2023enabling}.
While urban digital twins have been extensively studied in the literature, there also exists live urban digital twins from different parts of the world~\citep{ali2023enabling}: 
\begin{itemize}
    \item 
    Virtual Singapore\footnote{\url{https://www.nrf.gov.sg/programmes/virtual-singapore}} provides a detailed 3D model of the city, including texture, building models and more. Its main function is to support virtual experiments and test-bedding, such as analyzing network coverage and simulating crowd dispersion. Additionally, it plays a crucial role in urban planning, enabling the analysis of transport flow and pedestrian movement patterns.

   \item 
   Zurich, Switzerland, offers a leading-edge digital twin \citep{switerlandDT} employing sensors and drones to create a virtual representation of the city. 
   The primary focus is on urban design scenarios, offering planners, architects, and engineers a platform to assess and visualize the potential effects of proposed changes. 
   A notable attribute is its ability to simulate urban climate scenarios, analyzing interactions between the built environment and urban planning.

   \item Boston, USA, has developed a digital twin~\citep{bostonDT} incorporating diverse data such as buildings, transit routes, and tree canopies. The focus is on providing analytical tools to assess shadow projections and ensure new buildings adhere to zoning codes, including height and density specifications. Moreover, the platform aids various decision-making tasks, including flood modeling and line-of-sight evaluation.

   \item New South Wales (NSW), Australia, digital twin\footnote{\url{https://nsw.digitaltwin.terria.io/}} digitally represents the state's physical and geo-spatial features through satellite imagery, LiDAR, and other data sources. Like other live digital twins, the NSW digital twin aids in urban planning and asset monitoring, including roads, bridges, and utilities. It also contributes to emergency management by simulating  natural disasters such as bush-fires or flood.
\end{itemize}

Building on the existing spatial digital twins, in this paper, we use Sydney as a case study to explore how digital twins can contribute to urban sustainability. By integrating data from diverse sources, we aim to gain insights into factors influencing city resilience and identify potential improvements. We also provide vivid visualization and automated insights to make complex information more accessible and actionable. Additionally, we explore the application of machine learning to generate predictive models, such as for traffic crash risk.
Through this case study, we hope to demonstrate how a data-driven approach can inform sustainable urban planning and decision-making.

%% file: Spatial_Digital_Twin.tex
\section{Our Urban Digital Twin}\label{sec:Digital_Twin}
Our urban digital twin is a system designed to present spatio-temporal data from diverse dynamic datasets on a map. Through the integration of these disparate datasets, the platform offers user-friendly  tools that facilitate in-depth exploration of the data, enabling the visualization of datasets both on a map and across a time series. While we provide details of some representative applications of our system in Section~\ref{sec:applications}, next we present the details of our system and the datasets used.

\begin{figure*}[t]
\includegraphics[width=\textwidth]{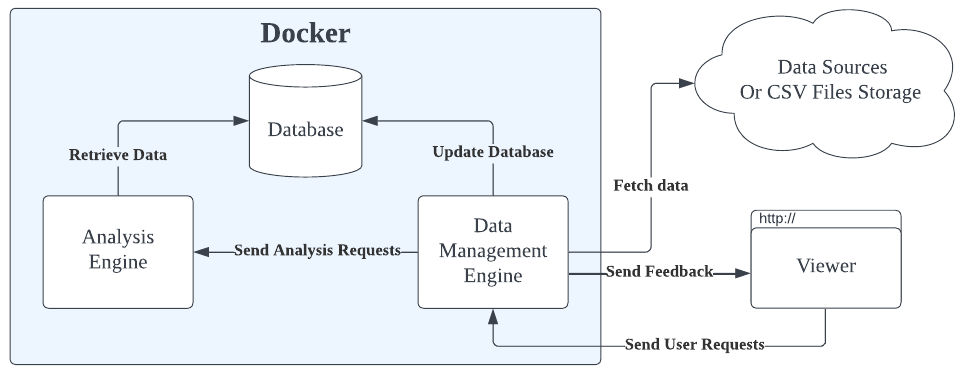}
\centering
\caption{A system diagram illustrating the workflow of our urban digital twin.}
\label{fig:system_chart}
\end{figure*}

\subsection{The System}


Our urban digital twin is composed of four essential components: a data management engine, an analysis engine, a database, and a viewer. Figure~\ref{fig:system_chart} shows the workflow of our system. Users interact with the interface (i.e., viewer), selecting the specific dataset/function they wish to visualize. The user requests are then finalized and sent to the data management engine. Depending on the nature of the requests, the data management engine may deconstruct them and invoke the analysis engine to obtain the results. Subsequently, the analysis engine retrieves the required data from the database, analyzes it, and provides feedback to the users. Additionally, to ensure the integration of newly released data, the data management engine is responsible for fetching data from the source dataset or CSV file storage and subsequently updating the databases. Next, we discuss the implementation details of each of these important components:

\begin{itemize} 

\item\textbf{Data Management Engine:}
The data management engine is responsible for gathering and storing data in the database. It consolidates data and delivers it to the frontend through a RESTful API (i.e., an interface that two computer systems use to exchange information securely over the internet).
The server is a Node.js\footnote{\url{https://nodejs.org/en}} application using TypeScript and the Express web framework. 
It primarily collects data through two methods: API retrieval and manual CSV uploads. For each new dataset, it includes an integration that manages the data's format and maps it to the database's table structure. It also fetches GIS information for point data and stores it in the database, enabling data aggregation at suburb and state levels.

\item\textbf{Analysis Engine:}
The analysis engine employs various techniques to automatically generate insights from the collected data. It is a Python application that uses various libraries such as GeoPandas\footnote{\url{https://geopandas.org/en/stable}} and Folium\footnote{\url{https://github.com/python-visualization/folium}} to generate correlation-based insights for each pair of columns in the underlying data and rank them based on their scores (details given in the next section). 

\item\textbf{Database:}
The database houses all the data in a normalized format and stores configurations for updating these datasets. It employs PostgreSQL\footnote{\url{https://www.postgresql.org}} and interfaces with both the server and the analysis engine.

\item\textbf{Viewer:}
The frontend facilitates dataset exploration, making requests to the server and the correlation engine, and visualizing the results. The frontend is a React.js\footnote{\url{https://react.dev}} application written in TypeScript. It utilizes Leaflet for data visualization on a mapping layer and Chart.js\footnote{\url{https://www.chartjs.org}} to represent time series data.

\end{itemize}

To facilitate seamless deployment and ensure consistency across various environments, all backend components (i.e., data managment engine, analysis engine and database) are maintained within a Docker container. 

\subsection{Datasets}
Our urban digital twin integrates a variety of datasets, including traffic, air pollution, crime, and more. Details of these datasets, including their sources and duration of availability, are presented in Table~\ref{table:data}. Specifically, each dataset is typically presented in a tabular structure, with records categorized into multiple aspects (e.g., the GHG emissions dataset encompasses emission data for various categories such as electricity, gas, waste, and wastewater). Each category includes associated GIS data and numerical measurements tracked over time.
The GIS data is represented as points or boundaries on earth surface. Through GIS mapping techniques, connections between points and boundaries are established, facilitating comparisons between measurements for point-based objects (e.g., traffic incidents) and boundary-based objects (e.g., crimes in a suburb).

\begin{savenotes}
\begingroup
\begin{table*}[t]
\centering 
\small
\begin{tabular} {c|c|c|c|c|c}
    \toprule
\textbf{Dataset} & \textbf{Freq.} & \textbf{Method}  & \textbf{Type} & \textbf{Source}& \textbf{Available Data} \\
 \midrule
 \midrule
Air Quality  & Daily& API & Point & NSW Gov\footnote{\url{https://www.dpie.nsw.gov.au/air-quality}} &  2019 - now \\
Traffic Volume  & Daily & API & Point & Transport NSW\footnote{\url{https://roads-waterways.transport.nsw.gov.au}} & 2019 - now\\
Traffic Incidents  & Second & API & Point & Transport NSW\footnote{\url{https://opendata.transport.nsw.gov.au}} & 2018 - now \\
Weather  & Hourly & API & Point & OpenWeatherMap\footnote{\url{https://openweathermap.org/}}  & 2018 - now\\
GHG Emissions & Annual & CSV & Boundary & City of Sydney\footnote{\url{http://data.cityofsydney.nsw.gov.au}}  & 2005 - 2018\\
Crime Incidents & Annual & CSV & Boundary & BOCSCAR\footnote{\url{http://www.bocsar.nsw.gov.au}} & 1995 - 2021 \\
  \bottomrule
\bottomrule
\end{tabular}
\caption{
The datasets integrated into our urban digital twins are obtained from various sources. We present information on each dataset, including its type, update frequency, access method, GIS type, data source, and the duration of the data available in our digital twin.}
\label{table:data}
\end{table*}
\endgroup
\end{savenotes}

Moreover, each dataset varies in its update frequency. For instance, the air quality and traffic volume datasets are updated on a daily basis, while the traffic incident and weather datasets receive updates every second and hourly, respectively. Consequently, these datasets are categorized into one of two types:

\begin{table}[t]
    \begin{tabular}{p{3.5cm}|p{12cm}}
        \toprule
        \textbf{Dataset} & \textbf{Attributes} \\
        \midrule
        \midrule
        Air Quality & CO, NEPH, NO, NO2, OZONE, SO2 \\
        \hline
        Weather & Temperature, Cloud, Gust, Dew Point, Pressure, Rain, Relative Humidity, Wind Speed, Wind Direction \\
        \hline
        GHG Emissions & Electricity, Gas, Transport, Waste, Waste Water, Other \\
        \hline
        Traffic Incidents & Heavy Traffic, Crash, Changed Traffic Conditions, Burst Water Main, Adverse Weather, Flooding, Traffic Lights Blacked Out, Breakdown, Scheduled Roadwork, Hazard, Traffic Lights Flashing Yellow, Special Event, Emergency Roadwork, Building Fire, Heavy Holiday Traffic, Ferry Out of Service, Roadwork, Late Finishing Roadwork, Traffic Signals, Accident, Bushfire, Grass Fire, House Fire, Fire, Hazard Reduction Burn, Special Event Clearways, Fog, Smoke, Hail, Bush Fire, Back-Burning Operation \\
        \hline
        Crimes & Abduction and kidnapping, Against justice procedures, Arson, Assault, Betting and gaming offences, Blackmail and extortion, Disorderly conduct, Drug offences, Homicide, Intimidation, stalking and harassment, Liquor offences, Malicious damage to property, Other offences, Other offences against the person, Pornography offences, Prohibited and regulated weapons offences, Prostitution offences, Robbery, Sexual offences, Theft, Transport regulatory offences \\
        \hline
        Traffic Volume & Count only \\
          \bottomrule
\bottomrule
    \end{tabular}
    \caption{Attributes in each dataset}
    \label{tab:data_attributes}
\end{table}

\begin{itemize}
\item Live datasets: These feature a queryable API, facilitating automatic retrieval of the data from these sources.
\item Published CSV datasets: These are released periodically by a designated data source and are manually uploaded to our digital twin.
\end{itemize}

In our system, priority is given to datasets with APIs, allowing for real-time analysis of new data as it becomes available. Nevertheless, the system also accommodates the manual upload of CSV files when necessary. 
When integrating data from disparate sources, the initial dataset may exhibit inconsistencies. To address this, the system automatically performs data cleansing procedures, such as averaging readings, mapping incident counts, removing duplicates, and extracting essential weather and air quality features. Table~\ref{tab:data_attributes} shows the attributes in each of the datasets.

%% file: Techniques.tex
\section{Representative  Sustainability Applications}\label{sec:applications}

This section delves into the sustainability applications of our urban digital twin. Specifically, we explore the following applications in this case study:

\begin{itemize}
\item \textbf{Interactive suburb ranking:} Users can explore rankings of suburbs based on chosen data and timeframes (Section~\ref{sec:suburbranking}). We showcase suburb rankings based on different types of emission data in the user-selected timeframe.
\item \textbf{Automated correlation insights:}  In Section~\ref{sec:stat-correlation}, we propose techniques to uncover hidden connections in data, revealing potentially interesting relationships, like traffic crashes and air pollution.
\item \textbf{Spatial correlation analysis:} In Section~\ref{sec:spatial-correlation}, we explore spatial connections
which  helps understand how factors like crime and emissions vary across locations.
\item \textbf{Traffic risk prediction:} We use our digital twin to predict traffic risk based on environmental data (Section~\ref{sec:Traffic_Analysis}).
\end{itemize}


\subsection{Interactive Suburb Rankings}\label{sec:suburbranking}

\begin{figure*}[t]
\includegraphics[width=\textwidth]{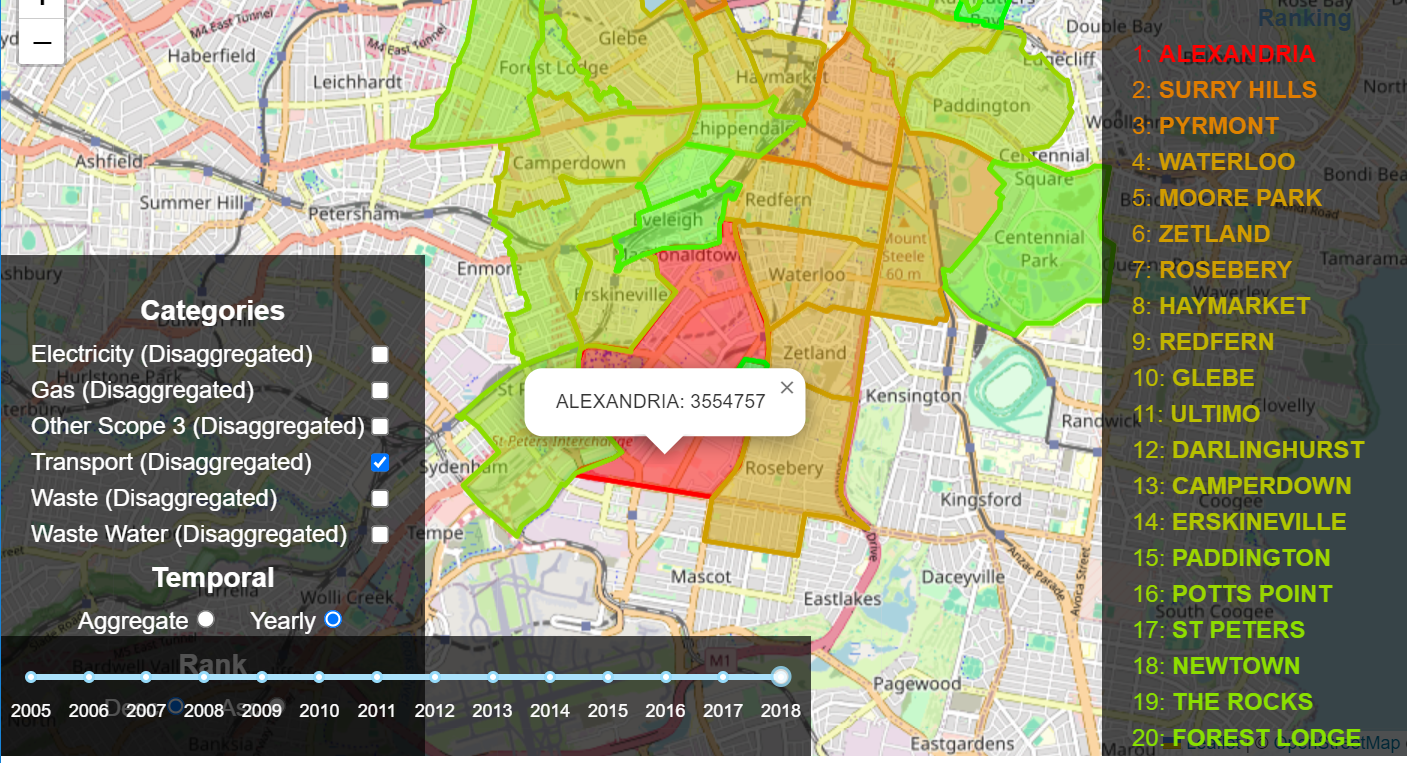}
\centering
\caption{Ranking of suburbs in descending order of transport-related emissions in 2018. Actual emission values can be seen by clicking on a suburb in the map. The Digital Twin allows visualizing the data sets according to various categories and across different years.  }
\label{fig:t_emissions}
\end{figure*}

We create an interactive tool to rank and visualize suburbs based on the underlying data considering the user-defined temporal range.  Figure~\ref{fig:t_emissions} presents a screenshot of our urban digital twin.  With our system, users can easily visualize the ranking of various suburbs based on user-selected input criteria. In the illustrated example, the figure displays suburb rankings in descending order of transport emissions, depicted from red to green, within the greenhouse gas emissions dataset for the specified year of 2018. Our visualization incorporates geographical elements, enabling users to easily identify the suburb with the highest emissions and assess the emission status of nearby suburbs. Furthermore, the visualization can be customized to rank suburbs by aggregating values across multiple categories, enhancing flexibility and depth in data exploration, e.g., rank suburbs based on emissions contributed by transport and electricity usage.

\input{Automated_Insights}

\input{Traffic_Risk_Analysis}

%% file: Automated_Insights.tex

\subsection{Automated Correlation Insights}\label{sec:stat-correlation}

In exploratory data analysis, insights are valuable and unexpected understandings derived from raw data. Various automated insights, including correlation-based insights, outlier detection, and trend analysis, have been proposed in the literature \citep{wongsuphasawat2016towards,demiralp2017foresight,srinivasan2018augmenting}. 
These insights, when combined with hypotheses or educated guesses, yield meaningful and actionable information, empowering stakeholders to make informed decisions. 
In this section, we explain how our urban digital twin can be used to uncover potentially interesting insights using the correlation coefficient.


We employ the \emph{Pearson Correlation Coefficient} to assess the correlation between two columns (or attributes) $c_i$ and $c_j$, although alternative correlations such as Kendall or Spearman could also be applied. The correlation coefficient between these columns is represented as $\rho(c_i, c_j)$, and the associated p-value is denoted as $p_{val}(c_i, c_j)$. In this context, the null hypothesis typically assumes no correlation between the variables. A lower $p_{val}$ indicates a more statistically significant observed correlation. We classify a correlation as strong if $\rho(c_i, c_j)>0.5$ (strong positive) or $\rho(c_i, c_j)<-0.5$ (strong negative). In other words, a correlation is considered strong if $|\rho(c_i, c_j)|>0.5$, where $|x|$ denotes the absolute value of $x$.

Let $C$ represent the set of columns across all datasets in the digital twin. The objective is to identify the k most \emph{interesting} correlation-based insights. One approach is to evaluate each pair of columns $c_i$ and $c_j$ and select the top-$k$ pairs with the highest $|\rho(c_i, c_j)|$.  However, a strong correlation may not necessarily be interesting for the following reasons: (i) a high correlation-coefficient between a pair of columns does not guarantee that each column is significantly important or aligns with the user's interests; (ii) some pairs with the highest correlation values may represent trivial insights, diminishing their importance to the user, e.g., we observed a strong positive correlation between two variables, ``rain'' and ``relative humidity'', which is clearly not interesting; and iii) merely focusing on the correlation coefficient is insufficient; it is essential to also factor in the p-value, as it acts as an indicator of the statistical significance of the correlation.

\subsubsection{Scoring The Insights}

To address the issues mentioned above, we propose techniques to score each insight (i.e., pair of columns) taking into account the importance of different columns and penalizing pairs of columns with uninteresting correlations.  
Specifically, to ensure that each column of an insight is important and meaningful, we define the impact score of a column $c_i$ as follow:
\begin{equation}
    impact(c_i) = \frac{card(\{|\rho(c_i, c_j)| > 0.5 | \forall c_j \in C \setminus c_i \} )}{card(C)}
\end{equation}

\noindent
where $card(S)$ denotes the cardinality of a set $S$.  The impact score measures the significance of a column within the context of the entire dataset. 
The numerator of the above equation computes, for a given column $c_i$, the number of other columns $c_j$ with which this column $c_i$ has a strong correlation (i.e., $|\rho(c_i, c_j)|> 0.5$). The impact score is subsequently determined by dividing this number by the total number of columns $card(C)$. 
Based on the impact score, we define the scoring function $S(c_i, c_j)$ of a pair of columns $c_i$ and $c_j$ as follow: 

\begin{equation}
\label{eq::score}
    S(c_i, c_j) = ( impact(c_i) + impact(c_j)) \times (1 - p_{val}(c_i,c_j))
\end{equation}

The scoring function evaluates and ranks correlated insights $(c_i, c_j)$ by combining the cumulative impact scores $impact(c_i)$ and $impact(c_i)$. Additionally, we incorporate the p-value $p_{val}(c_i, c_j)$ to ensure that the observed correlation is not a mere outcome of random chance. Since a lower $p_{val}$ signifies a more statistically significant observed correlation, we utilize the value of $(1 - p_{val}(c_i, c_j))$.
To further improve the extraction of meaningful insights, we  apply a penalty within the scoring function for the column pairs belonging to the same category, e.g., a penalty is applied if both columns are from the weather dataset. This penalty mitigates the inclusion of trivial correlations, such as rain and relative humidity or rain and temperature.

Note that different datasets are obtained from different sources and may have different granularity, e.g., air quality data readings are daily whereas weather data is hourly. In such cases, we consistently aggregate the values of the higher frequency column into the lower frequency column. For example, when dealing with two columns from the air quality and weather, we aggregate the hourly weather data into daily data, e.g., total rain in the day, average temperature on the day etc. Additionally, we employ binning or bucketing to discretize continuous data into distinct groups.

\subsubsection{Recommending Interesting Insights}

In order to recommend the most interesting correlation-based insights, our urban digital twin computes the absolute correlation for every column pair, eliminating those with absolute value of correlation coefficient less than 0.5. For the remaining column pairs, our system calculates their ranking score using Equation~\eqref{eq::score}. After applying the penalty, our system  returns the top-$k$ correlation-based insights based on their adjusted ranking scores.

\begin{figure*}[t]
\includegraphics[width=\textwidth]{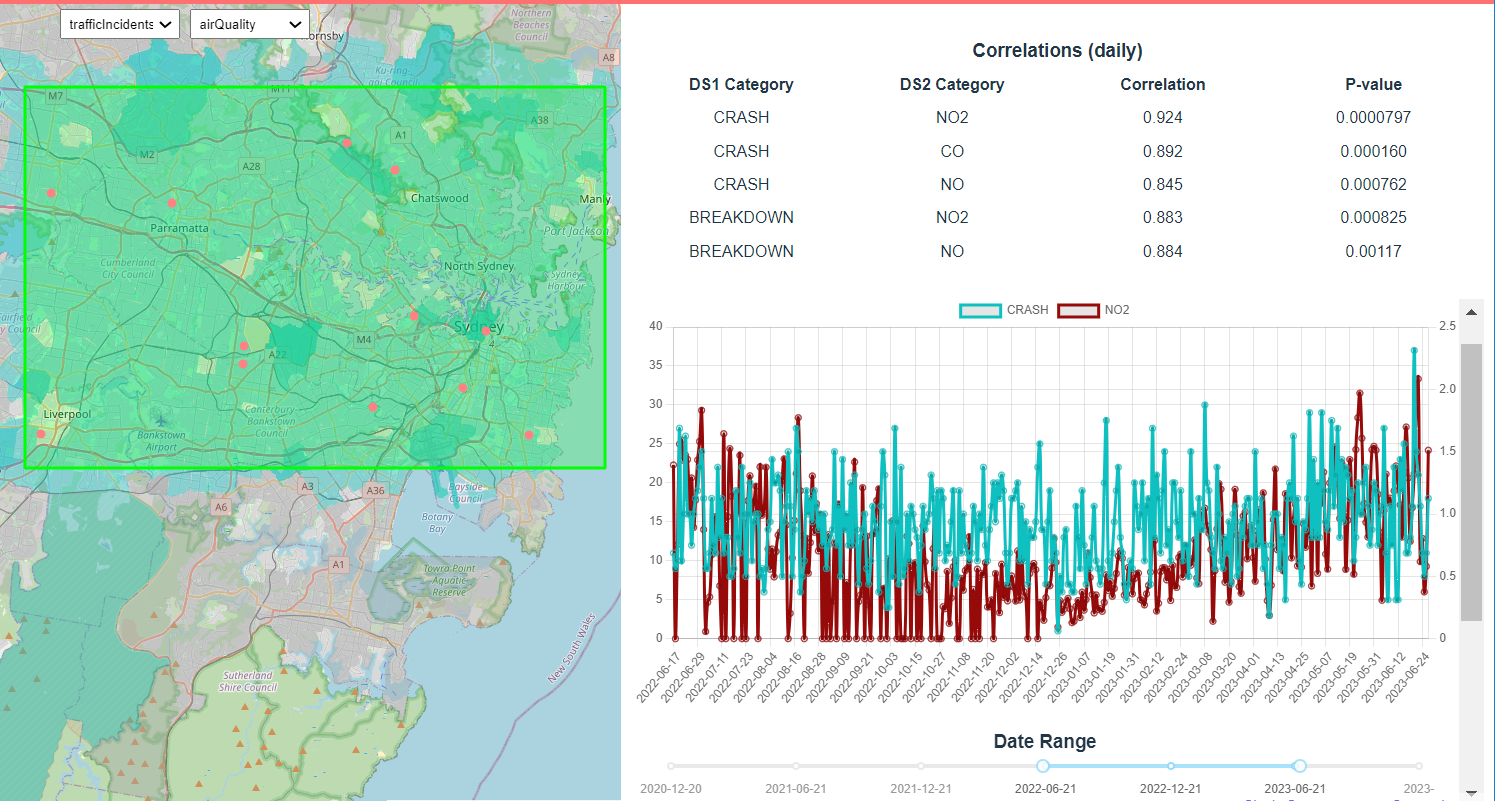}
\centering
\caption{The digital twin finds the most prominent correlations from underlying data considering the user-selected spatial and temporal ranges. Here, the user-selected area is shown as a green shaded rectangle and the temporal range is 21-Jun-2022 to 21-Jun-2023. Five most prominent correlations are shown and the chart visualises traffic crashes and NO2. 
}
\label{fig:pairwise_correlation}
\end{figure*}

Figure~\ref{fig:pairwise_correlation} illustrates an example of our urban digital twin for recommending statistically correlated insights. On the left-hand side, our system displays a geographical visualization of Sydney metropolitan area. The users can draw a rectangular area (highlighted in green) on the map to select a region of interest. The users can also select a temporal range. In this example, the user-selected temporal range is 21-June-2022 to 21-June-2023.
Upon selecting the spatial and temporal ranges, our urban digital twin automatically returns some prominent statistically correlated insights, along with their corresponding correlation coefficients and p-values.
In Figure~\ref{fig:pairwise_correlation} (see top right), for the selected spatial and temporal ranges,  our system identifies positive correlations between traffic crashes/breakdown and air pollution such as NO, CO and NO2. While some existing research has shown that air pollution may increase the risk of crashes~\citep{sager2019estimating}, we emphasize that correlation does not imply causation. Therefore, these insights should be seen as potentially interesting observations in the data that need to be carefully investigated further before reaching any conclusions.
In Figure~\ref{fig:pairwise_correlation}, the bottom right displays visualizations for the top-ranked insight between crashes and NO2 (shown in blue and red, respectively).

\subsection{Spatial Correlation Analysis}
\label{sec:spatial-correlation}

In the previous subsection, we explore the correlation coefficient based insights, which is based on the statistical relationship between two variables, indicating how changes of values in one variable affect the other.
Spatial autocorrelation~\citep{getis2009spatial}, on the other hand, measures the correlation of a variable across space, considering relationships with neighboring data points. This can result in positive correlations (similar or clustered values), neutral correlations (no specific pattern), or negative correlations (dissimilar or dispersed values). The positive correlations give an important insight of how a particular variable, e.g., green house gas emissions, are clustered in different regions. 

In this section, we explore how our urban digital twin can be  used to compute spatial autocorrelation to find interesting spatial insights of different variables. There are two types of spatial autocorrelation: global spatial autocorrelation and local spatial autocorrelation~\citep{HAINING2001}.

\begin{table*}[t]
\centering
\begin{tabular}{c|c|c}
    \toprule
Dataset   & Variable   & Global Autocorrelation  \\
 \midrule
 \midrule
Crimes Incidents  & All Crimes       & 0.40                      \\
GHG Emissions & Waste Water        & 0.40                    \\
GHG Emissions    & Other Scope    & 0.37                  \\
GHG Emissions & Gas & 0.28                    \\
GHG Emissions & Electricity        & 0.27                   \\
GHG Emissions & Waste      & 0.26                   \\
GHG Emissions & Transport     & 0.24                    \\
Air Quality & CO       & 0.17                 \\
  \bottomrule
\bottomrule
\end{tabular}
\caption{\label{tab:golbal_autocorrelation} Variable names and global auto-correlation values of top ranked variables (in decreasing order of their correlation values).}
\end{table*}




\subsubsection{Global Spatial Autocorrelation}
Global spatial autocorrelation assesses overall spatial clustering trends in a dataset, helping determine if there is a meaningful spatial pattern. For example, if the air quality in different suburbs are randomly distributed, no significant clustering of similar values would be observed on a map. 

To measure global autocorrealtion, a statistical measure named Global Moran's \emph{I}~\citep{anselin1995local} is generally used, which gives a formal indication of the degree of linear association between the value of a variable at a location and the weighted average of the neighborhood values. Let there be $n$ data points representing $n$ locations. Assume $x_i$ is the observed value of location $i$ and $x_j$ is the observed value of a neighboring location $j$ of $i$. We can compute the global Moran's \emph{I} index, a measure of the global spatial autocorrelation, as follows.

\begin{equation}
    \emph{I} = \frac{\sum_{i=1}^{n}\sum_{j=1}^{n}(x_i - \overline{x})(x_j - \overline{x})}{\sum_{i=1}^{n}\sum_{j=1}^{n}w_{ij}\sum_{i=1}^{n}(x_i - \overline{x})}
\end{equation}

\noindent
where, $w_{ij}$ is the spatial weight matrix (set 1 if $i$ and $j$ are neighbors, and 0 otherwise) determining the spatial relationship between location $i$ and $j$, and $\overline{x}$ is the mean value of the variable.

\begin{figure}[t]
\centering
	\subfigure[All Crimes]	{\centering \label{fig:cor:crimes}\includegraphics[width=0.49\columnwidth]{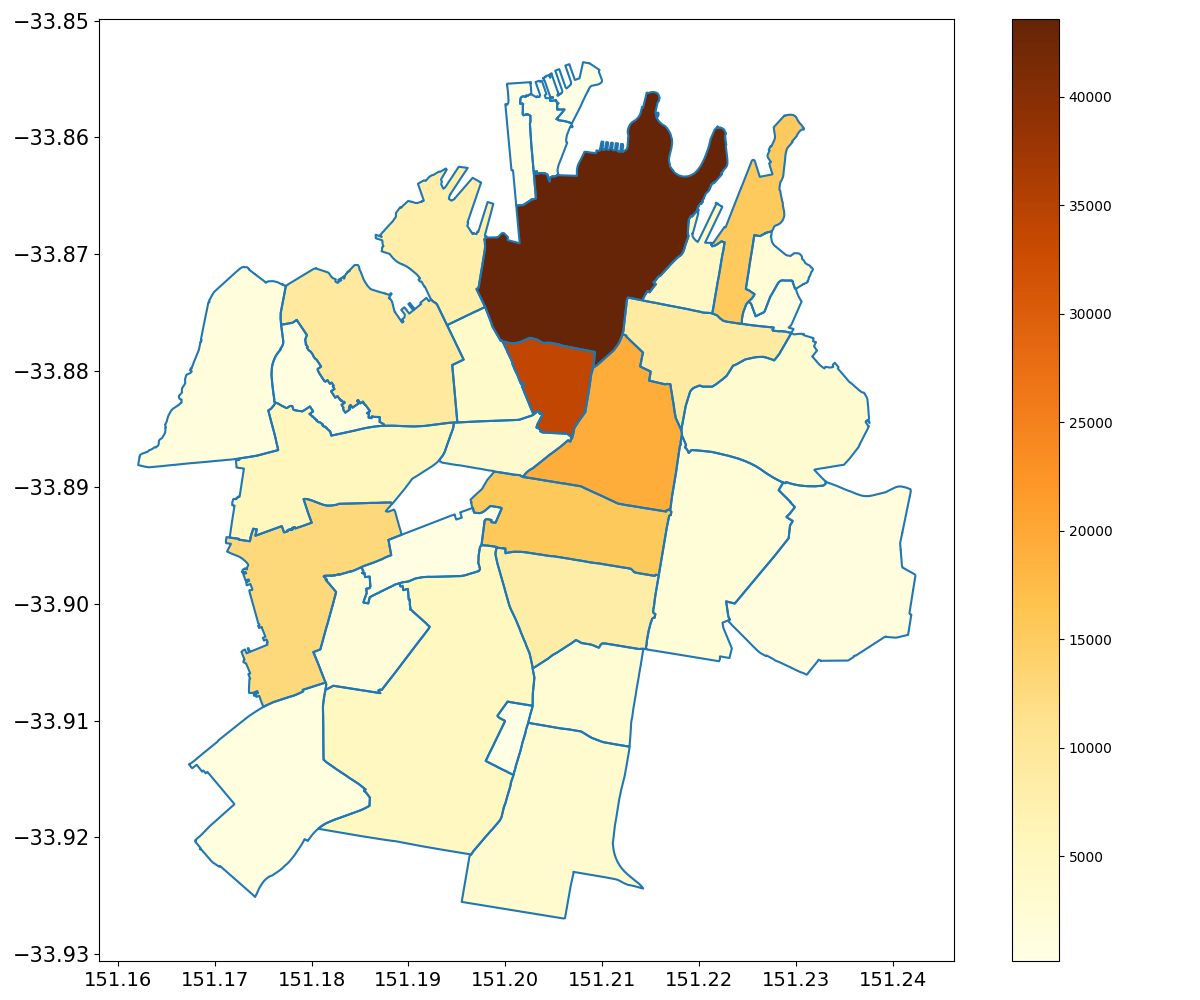}}
	\subfigure[Emissions (Waste water)]
{\label{fig:cor:emissions}\includegraphics[width=0.49\columnwidth]{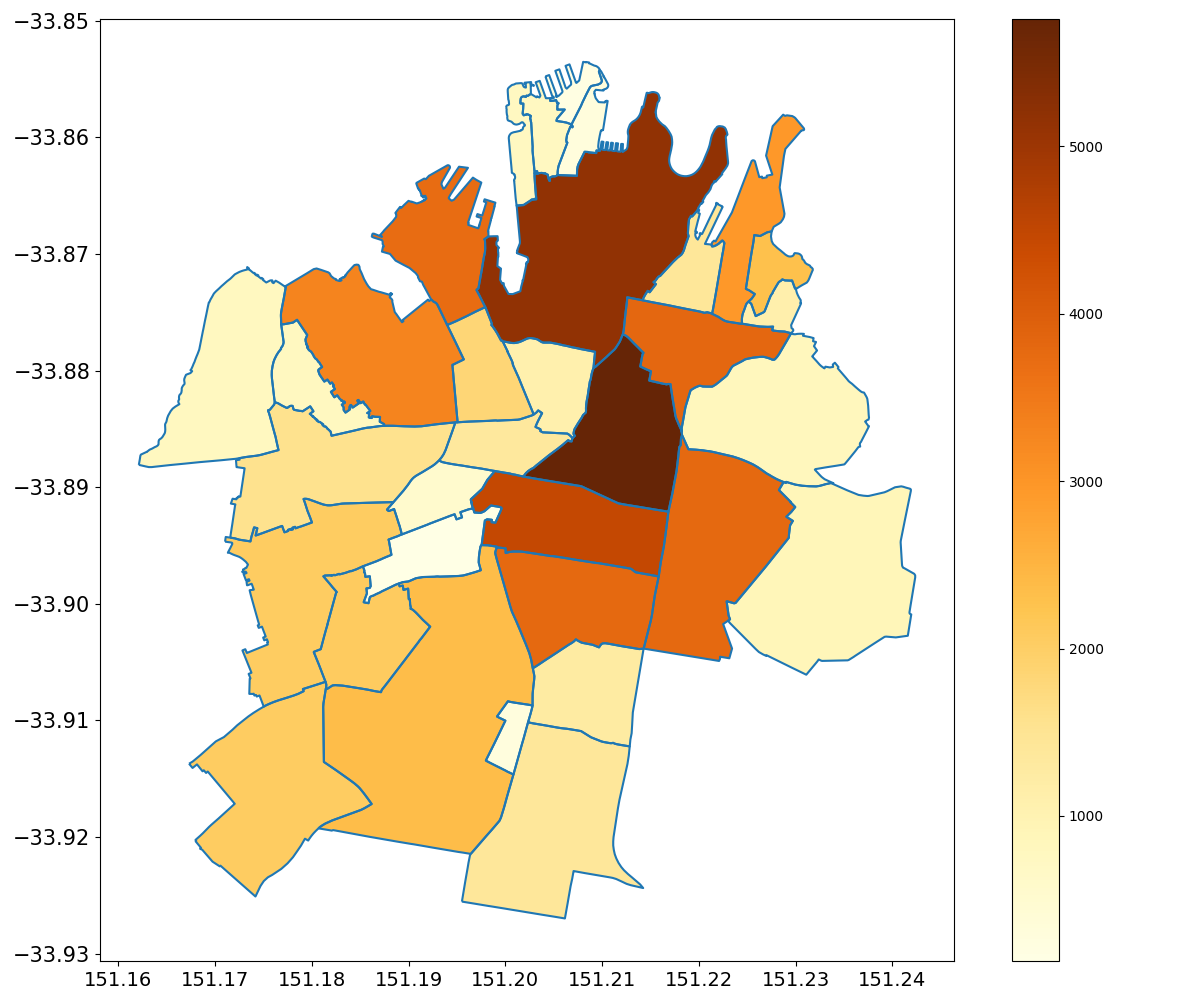}}
	\caption{Spatial correlation}
	\label{fig:spatial_correlations}
\end{figure}

We analyze the global spatial auto-correlation of different variables in our urban digital twins. The high value of $I$  means that our dataset is positively spatially correlated, i.e., spatial clusters are present. For example, we consider each value of a variable (e.g., number of crimes) in a suburb as a data point, and the suburbs that share their boundaries are considered as neighbors. There were $33$ suburbs in our Sydney dataset, in this example. Table~\ref{tab:golbal_autocorrelation} shows the top eight variables ranked based on global spatial autocorrelation values. We have observed that total number of crimes (regardless of their type) and waste water emissions are the top-2 spatially correlated variables. We can see in our spatial heatmaps of the two variables (where different range of values are represented using different colors) in Figure~\ref{fig:spatial_correlations} that there are some spatial clusters visible on the map.

\subsubsection{Local  Spatial Autocorrelation}
The global autocorrelation provides us with a single value for our entire dataset to assess whether the variable of interest has a spatial correlation. On the other hand, local spatial autocorrelation (through local Moran's I) identifies whether a particular location exhibits a spatial correlation with its neighbors. Essentially, the Local Moran's I statistic is a measure of the difference between observed value at location $i$ and the mean, multiplied by the sum of differences of its neighbors and the mean. We can compute the local Moran's $I_i$ index of location $i$ as follows.

\begin{equation}
    \emph{I*} = \frac{(x_i - \overline{X})}{S_i}\sum_{j=1}^{n}(x_j - \overline{X})
\end{equation}

\noindent
where, $S_i$ is the standard deviation.


\begin{figure*}[t]
\includegraphics[width=0.6\textwidth]{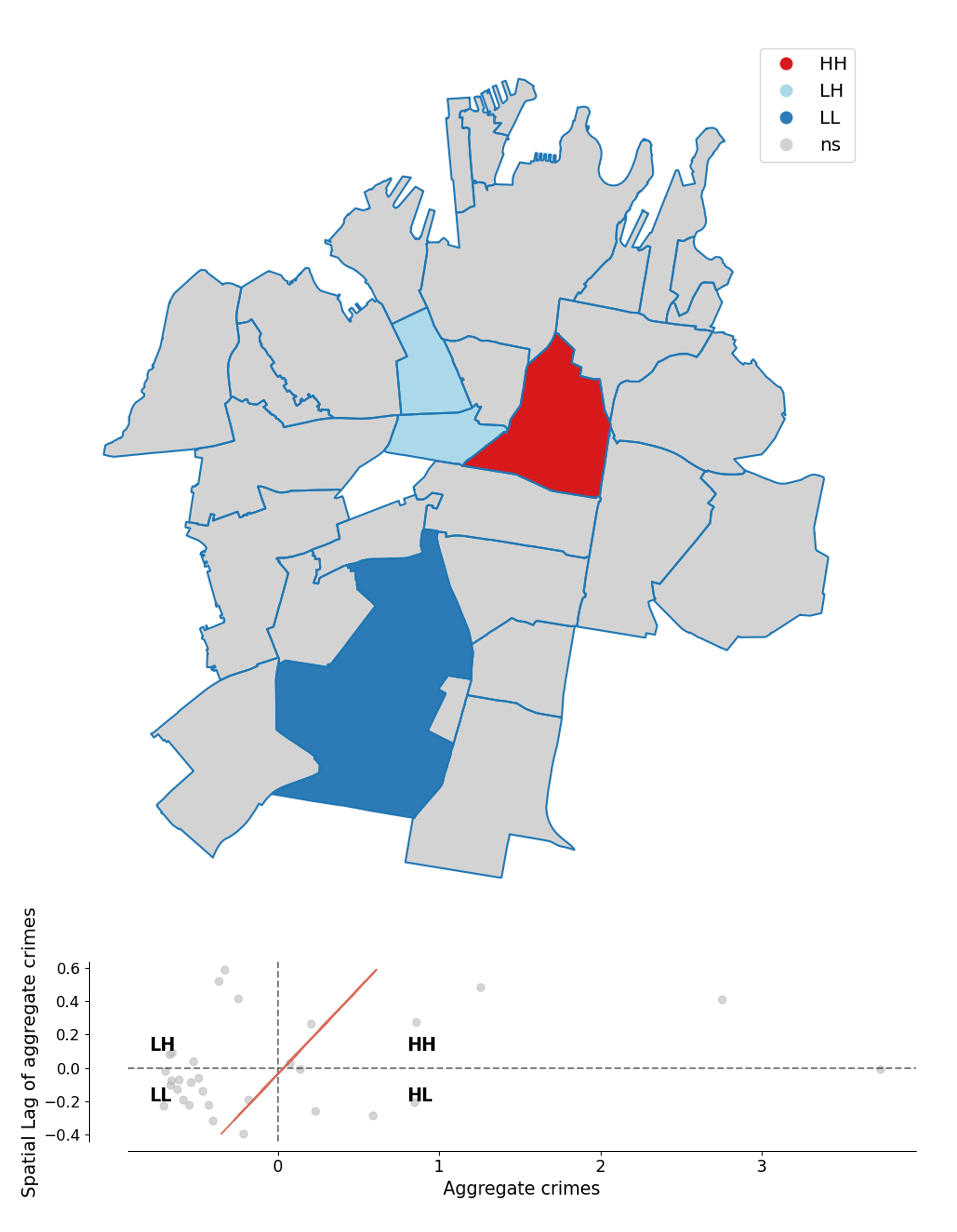}
\centering
\caption{Local autocorrelaton maps (aka Spatial Lag Choropleth Map) with Moran local scatter plot of Crimes}
\label{fig:slag_crimes}
\end{figure*}

\begin{figure*}[t]
\includegraphics[width=0.6\textwidth]{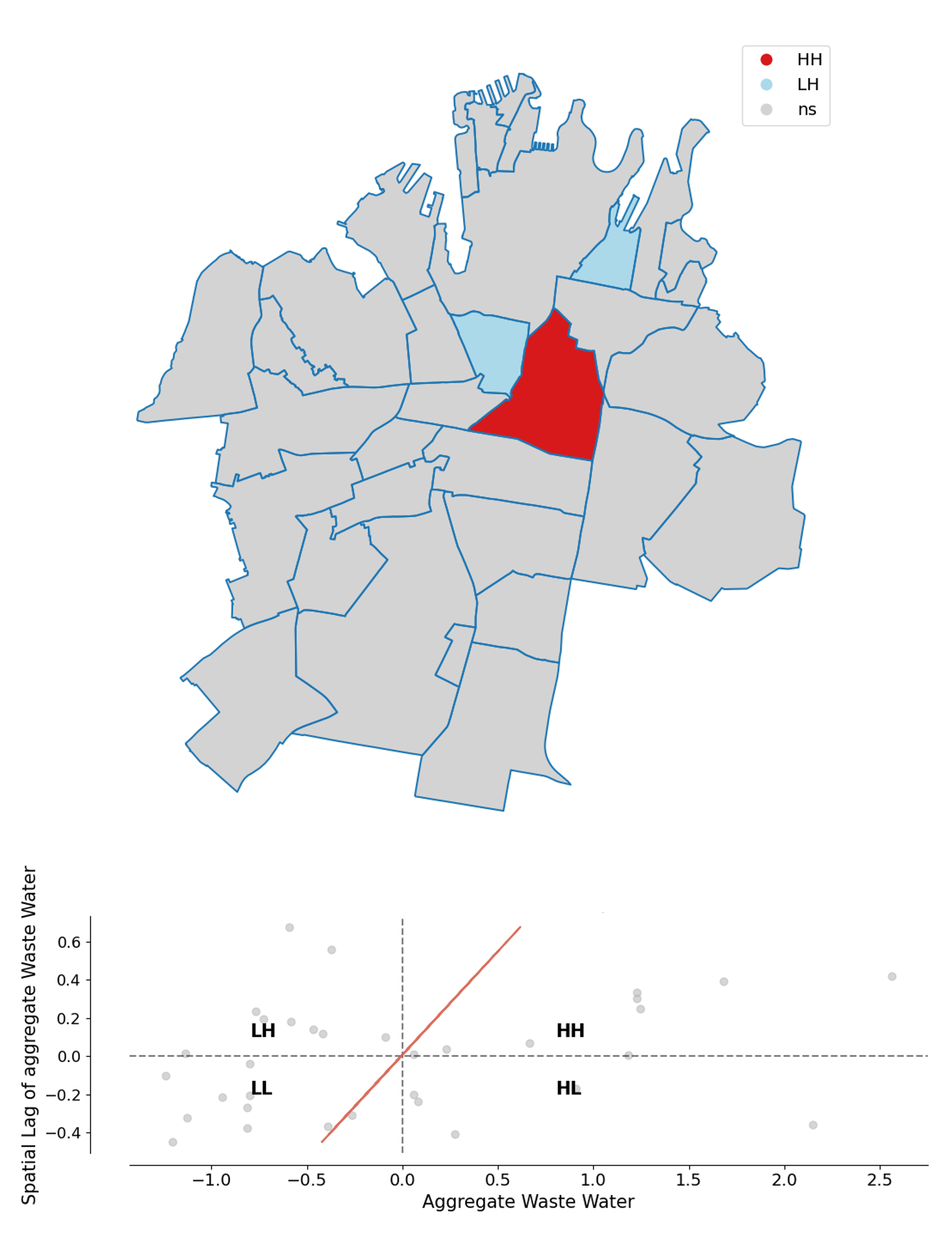}
\centering
\caption{Local autocorrelaton maps (aka Spatial Lag Choropleth Map) with Moran local scatter plot of emissions}
\label{fig:slag_emissions}
\end{figure*}

Based on local Moran's values, we can categorize spatial clusters into four types: HH (high values near high), LL (low values near low), LH (low values near high), and HL (high values near low) using Local Indicators of Spatial Association (LISA)~\citep{anselin1995local}. Intuitively, LISA essentially arranges spatial clusters in the above mentioned four categories.  

We have run local Moran's analysis for the two variables with high global spatial autocorrelation (top-2) as observed in Table~\ref{tab:golbal_autocorrelation}. Figures~\ref{fig:slag_crimes} and~\ref{fig:slag_emissions}  show the local autocorrelaton maps (aka Spatial Lag Choropleth Map) with Moran local scatter plot (at the bottom) for Crimes and Emissions (Waste Water), respectively, for Sydney suburbs. In  Figure~\ref{fig:slag_crimes}, the cell shown in red is a high-high cluster, i.e., this suburb (Surry Hills) and its neighboring suburbs have high crime rate (larger than mean), the cells shown in light blue represent a low-high cluster, i.e., these suburbs (Ultimo and Chippendale) have a low crime rate whereas their neighboring suburbs have high crime rates, and the cell shown in dark blue (Alexandria) is a low-low cluster, where this suburb and its neighboring suburbs have low crime rates. If a cluster is deemed `not significant' (displayed as `ns'), it means that the observed pattern of values in that location is not different enough from what would be expected under the assumption of spatial randomness. Therefore, the cells shown in gray color represent `not significant' clusters.  Likewise, we can explain the spatial clusters of Emissions (Waste Water) from Figure~\ref{fig:slag_emissions}. These clusters can help the urban/city planner to monitor location based clusters and act accordingly. At the bottom of Figures~\ref{fig:slag_crimes} and~\ref{fig:slag_emissions}, moran local scatterplot shows some degree of positive spatial autocorrelation. For this, spatial lag is computed which involves considering the assigned weights for each neighboring unit and multiplying them by the corresponding variable values. Subsequently, it calculates the sum of these weighted values across all neighbors. Essentially, it represents the ``local average'' of a variable for each observation, taking the neighborhood into account.

%% file: Traffic_Risk_Analysis.tex
\subsection{Traffic Risk Prediction}\label{sec:Traffic_Analysis}

Road crashes, as highlighted in a 2018 World Health Organization (WHO) report, rank among the foremost causes of death and severe injuries, claiming over 1.35 million lives and causing up to 50 million injuries annually, with an associated economic cost of 518 billion dollars per year and inflicting lasting physical disabilities and mental trauma on numerous families~\cite{sohail2023data}.
To establish safer and more efficient road networks, predicting the risk of traffic accidents is crucial for preventing their occurrence, issuing risk alerts, and proactively minimizing damages.
In this section, we utilize data from our urban digital twin and apply machine learning techniques to investigate and forecast traffic risks under diverse environmental conditions.



\subsubsection{Data Preprocessing}
Data, covering air quality,  weather, and traffic incidents, is obtained from the urburn digital twin database. Within these datasets,  certain features exhibited high correlations, such as ozone, NO, NO2 from air quality, humidity, cloud cover, and minimum and maximum temperature from weather. To enhance interpretability and prevent redundancy, only features with the highest correlation to the number of crashes are utilized for analysis. Consequently, the final dataset includes suburbId, \#crash, CO, NO, SO2, NEPH, humidity, precipitation, temperature, pressure, wind speed, and wind direction.

Following common machine learning practices, the dataset is split into training and testing sets with an 80:20 stratified split. The training set consists of 15,328 samples, with 12,743 classified as low-risk (class 0, indicating no incident) and 2,585 as high-risk (class 1, indicating at least one incident occurred). The test set comprises 3,832 samples, with 3,208 from low risk and 624 from high-risk samples.

This dataset shows the problem of high class imbalance where low-risk class dominates the high-risk class. To address this, we implemented the Synthetic Minority Over-sampling Technique combined with Tomek links (SMOTE-Tomek)~\citep{SMOTE-Tomek} using Python imbalanced learn library\footnote{\url{https://pypi.org/project/imbalanced-learn/}}. This approach is a widely used oversampling technique that generates synthetic instances for the minority class by interpolating between existing instances.
Note that, we also explored other under or oversampling techniques, such as random under-sampling~\citep{random_us}, NearMiss~\citep{nearmiss}, Edited Nearest Neighbors~\citep{ENN}, Adaptive Synthetic Sampling~\citep{ADASYN}, among others. However, we present the results only for SMOTE-Tomek because of its better results overall. After oversampling, the training dataset contains 12,527 samples for both low-risk and high-risk classes.

\subsubsection{Traffic Risk Prediction with Environment}

To accurately  predict traffic risks based on diverse environmental data, machine learning algorithms are preferred over deep learning due to the limited size of the dataset. 
Specifically, this study evaluates the performance of several popular machine learning algorithms, including Decision Tree~\citep{Decision_Tree}, Naive Bayes~\citep{naive_bay}, Logistic Regression~\citep{logistic}, Gradient Boosted Tree~\citep{GBDecisionTree}, and Random Forest~\citep{random_forest}.
All machine learning models are implemented using the Python Sci-Kit Learn library\footnote{\url{https://scikit-learn.org/stable/}} and are fine-tuned utilizing the HyperOpt library\footnote{\url{https://hyperopt.github.io/hyperopt/}}.
For the hyperparameter tuning process using HyperOpt, it is essential to define a search space for the parameters to be optimized. The objective function aims to identify the optimal set of hyperparameters by maximizing the negative value of the area under the curve (AUC). AUC is chosen as the performance metric over F1-score or accuracy because it is more robust to class distribution disparities of the imbalanced dataset~\cite{AUC}.

Table~\ref{tab:performance} presents the performance of these machine learning algorithms on the test dataset. The measurement includes various standard metrics such as prediction time in seconds, accuracy, weighted F1-score, precision, recall, and AUC. 
The results show that Gradient Boosted Tree (GBT) has the best performance in terms of precision and AUC whereas Randomg Forest (RF) outperforms the other algorithms in terms of accuracy, F1-score and recall. Overall, these algorithms have quite good performance across various evaluation metrics. The prediction speed of all algorithms is very fast but Decision Tree has the smallest prediction time. It is important to note that these promising results are obtained by considering only the weather-related data, and there is potential for further improvement in performance by integrating other types of datasets in the digital twin.

\begin{table*}[]
\centering
\begin{tabular}{c|c|c|c|c|c|c}

    \toprule
 \textbf{Method}       & \textbf{Time} & \textbf{Accuracy} & \textbf{F1-Score} & \textbf{Precision} & \textbf{Recall} & \textbf{AUC}\\
 \midrule
 \midrule
                                      \textbf{DT}           & \textbf{\underline{0.001}}                  & 0.718                          & 0.751                       & 0.825               & 0.718 & 0.705                 \\\hline
                                    \textbf{GBT}           & 0.009                  & 0.680                          & 0.721                      & \textbf{\underline{0.834}}                & 0.680 & \textbf{\underline{0.715}}                 \\\hline
                                    \textbf{LR}  & 0.002                  & 0.604                          & 0.655                        & 0.766                & 0.604    & 0.578              \\\hline
                                    \textbf{NB} & 0.002                  & 0.592                         & 0.645                        & 0.782                & 0.592 & 0.607                 \\\hline
                                    \textbf{RF}    & 0.064                  & \textbf{\underline{0.747}}                          & \textbf{\underline{0.768}}                      & 0.802                & \textbf{\underline{0.747}}    & 0.656               \\

  \bottomrule
\bottomrule
\end{tabular}
\caption{\label{tab:performance}Performance metrics comparison of machine learning models trained on the SMOTE-Tomek sampled dataset. Results are presented for diverse machine learning models, including Decision Tree (DT), Gradient Boosted Tree (GBT), Logistic Regression (LR), Naive Bayes (NB), and Random Forest (RF). The table highlights essential evaluation metrics, including prediction time (in seconds), accuracy, weighted F1-score, weighted precision, weighted recall, and Area Under the Curve (AUC). The best value for each evaluation metric is shown in bold and is underlined.}
\end{table*}

Overall, the results indicate that by choosing an appropriate machine learning algorithm, the data sources from our urban digital twin can be leveraged for accurate prediction of traffic risks, providing an effective means to proactively prevent traffic incidents.
For example, using our digital twins, one can predict traffic risks for different suburbs, allowing for the allocation of more workforce or deploying other strategies to prevent accidents in high-risk areas.
While this study focuses on standard machine learning techniques, further improvement can be achieved by incorporating advanced techniques for feature selection, addressing class imbalance, and more.
Future work will contribute to the continual advancement and practical implementation of machine learning in traffic risk prediction for urban digital twin.